\documentclass[]{interact}

\usepackage[normalem]{ulem}
\usepackage{xcolor}
\usepackage{comment}
\usepackage{lmodern}
\usepackage[authoryear]{natbib}
\usepackage{epstopdf}
\usepackage[caption=false]{subfig}
\usepackage{enumitem}

\theoremstyle{plain}
\newtheorem{theorem}{Theorem}

\theoremstyle{definition}

\theoremstyle{remark}

\begin{document}

\title{Nonparametric Estimation under General Nonlinear ODE Constraints: A Comparison with Parametric ODE-Fitting Methods}

\author{
\name{Chunlei Ge\textsuperscript{a} and W. John Braun$^\ast$\textsuperscript{a}\thanks{$^\ast$CONTACT W. John Braun. Email: john.braun@ubc.ca}}
\affil{\textsuperscript{a}The University of British Columbia Okanagan, 3333 University Way, Kelowna, BC, Canada}
}

\maketitle

\begin{abstract}
Many physical, biological, and epidemiological processes are governed by ordinary differential equations (ODEs) that are nonlinear in the state variable, including logistic population growth, chemical reaction kinetics, and epidemiological compartment models. We develop a differential equation-constrained local polynomial regression (DE-constrained LPR) framework for the general first-order ODE constraint g'(x) = F(x, g(x)), where F may be any Lipschitz continuous function, extending prior work restricted to exponential and linear ODE structures. Because F is generally nonlinear in g, the Taylor coefficients of the DE1-k estimator cannot be written in closed form; instead they are obtained by successive symbolic differentiation of F, and the estimator is computed by nonlinear least squares, requiring only a single local parameter at each evaluation point regardless of polynomial degree k. We derive the asymptotic conditional bias and variance of the DE1-k estimator, propose an AIMSE-optimal bandwidth that exploits the ODE structure to avoid direct estimation of high-order derivatives, and evaluate the method in a simulation study based on logistic growth, benchmarking against the parameter cascading method of Ramsay et al. (2007) (PCODE) and classical local linear regression. The DE-constrained estimator consistently outperforms local linear regression and is competitive with PCODE even though it estimates no structural parameter of the ODE; a sensitivity analysis across growth rates shows DE-constrained estimation becomes more accurate and more robust than PCODE as the curve steepens and PCODE's parameter estimation grows less stable. These results position DE-constrained LPR as a practical nonparametric alternative to parametric ODE-fitting methods when structural parameters are difficult to identify reliably.
\end{abstract}


\section{Introduction}
\label{sec:intro}

Many quantities of scientific interest evolve according to a known or partially known ordinary differential equation (ODE), yet are observed only through noisy, discretely sampled data. Logistic population growth, chemical reaction kinetics, and epidemiological compartment models are all governed by first-order ODEs that are nonlinear in the state variable, and in each case the applied goal is the same: to recover the underlying curve $g(x)$ nonparametrically while making full use of the mechanistic structure implied by the ODE, rather than either ignoring that structure (as in classical local polynomial regression) or committing to a fully parametric solution of the ODE (as in nonlinear least squares fitting of a closed-form or numerically integrated solution).

Differential equation-constrained local polynomial regression (DE-constrained LPR) addresses this problem by embedding the ODE directly into the local polynomial fitting criterion, so that the Taylor expansion used to approximate $g$ near an evaluation point is built from the ODE itself rather than from unconstrained polynomial coefficients. This approach has been developed for two restricted classes of first-order ODE: the exponential growth model $g'(x) = \lambda g(x)$, applied to tumour growth data \citep{ge2026differential}, and the general linear model $g'(x) = a(x)g(x) + b(x)$, applied to firebrand burning-rate data in a companion paper. Both are special cases of the general nonlinear constraint
\[
g'(x) = F(x, g(x)),
\]
where $F$ may be any Lipschitz continuous function of $x$ and $g$. Many of the most important applied ODE models, including logistic growth, Michaelis--Menten kinetics, and the compartment models used in epidemiology, are nonlinear in $g$ and are not covered by the linear framework.

In this paper we develop DE-constrained LPR for the fully general nonlinear constraint $g'(x) = F(x, g(x))$. The key technical difficulty relative to the linear case is that the Taylor coefficients of $g$ at a target point can no longer be written in closed form as linear combinations of $g(x_0)$; instead, each higher-order derivative is obtained by successive symbolic differentiation of $F$ via the chain rule, and the resulting local approximant is nonlinear in the single unknown $\alpha = g(x_0)$. The DE1-$k$ estimator is therefore defined as a nonlinear least squares (NLS) problem rather than the closed-form weighted least squares problem of the linear case, and is computed with a pilot estimator supplying warm-start values for the NLS optimizer. A central practical advantage carries over from the linear framework: because every higher-order Taylor coefficient is expressed in terms of the single unknown $\alpha = g(x_0)$ via the ODE, only one local parameter needs to be estimated at each evaluation point, regardless of the polynomial degree $k$, in contrast to classical local polynomial regression of degree $p$, which requires estimating $p+1$ local parameters.

A separate strand of statistics addresses ODE-constrained estimation through globally parametric or semiparametric means, most notably the parameter cascading (generalized profiling) framework of \citet{ramsay2007parameter}, implemented in the \texttt{pCODE} package \citep{wang2022pcode}. Parameter cascading represents the solution curve in a spline basis and estimates the structural parameters of the ODE by a nested optimization in which the spline coefficients are profiled out at an inner level while the structural parameters are estimated at an outer level. This approach can be highly accurate when the parametric form of the ODE is exactly correct and its structural parameters are reliably identifiable, but the outer optimization is generally nonconvex, and its performance can degrade, sometimes sharply, when the identification problem is difficult. DE-constrained LPR offers a complementary approach: it treats the ODE as a known constraint on the shape of $g$ rather than as a parametric model to be fitted, and so it estimates no structural parameter at all. This makes it a useful point of comparison for understanding what a purely nonparametric, structure-respecting estimator can achieve relative to a parametric ODE-fitting method, and where the two approaches diverge.

The remainder of the paper is organized as follows. Section~\ref{sec:model} defines the general nonlinear DE-constrained regression model and states the regularity assumptions used throughout. Section~\ref{sec:est} derives the DE1-$k$ estimator, including the symbolic construction of ODE-derived Taylor coefficients, the nonlinear least squares estimation step, and the pilot estimator used for warm-starting. Section~\ref{sec:theory} establishes the asymptotic conditional bias and variance of the DE1-$k$ estimator. Section~\ref{sec:bandwidth} develops an AIMSE-optimal bandwidth and a practical two-step bandwidth selector that exploits the ODE structure to avoid direct estimation of high-order derivatives. Section~\ref{sec:sim} reports a simulation study based on logistic population growth, comparing DE1-3 against PCODE and classical local linear regression, together with a sensitivity analysis across growth rates. Section~\ref{sec:discussion} discusses the implications and limitations of the results, and Section~\ref{sec:conclusions} concludes.

\section{General Nonlinear DE-constrained Regression Model}
\label{sec:model}

Given $n$ independent observations on an explanatory variable $x$ and a response variable $y$, we consider the general nonlinear DE-constrained regression model
\begin{equation}
y_i = g(x_i) + \varepsilon_i
\mbox{ where } g'(x) = F(x,g(x)), \quad x_i \in [a,b], \quad i = 1, 2, \ldots, n,
\label{equ:general}
\end{equation}
where $F$ may be any Lipschitz continuous function and $\varepsilon_i$ are uncorrelated, mean-zero errors.

We work under the following regularity assumptions.
\begin{enumerate}[label=(\Roman*)]
  \item The density $f(x)$ of the design points is continuous, bounded away from zero on
        $[a,b]$, and has a continuous derivative.
  \item The regression function $g(x)$ has $k+2$ continuous derivatives in a neighbourhood of $x_0$. For the constraint function $F$, the partial derivative $\frac{\partial^k F}{\partial^{k-1} x\,\partial g}$ is Lipschitz continuous.
  \item The kernel $K$ is a symmetric probability density function with
        $\mu_{k+1}(K) = \int u^{k+1} K(u)\,du$,
        $\mu_{k+2}(K) = \int u^{k+2} K(u)\,du$, and
        $R(K) = \int K^2(u)\,du$ all finite.
  \item The errors $\varepsilon_i$ are uncorrelated with mean zero and finite variance
        $\sigma^2 < \infty$.
\end{enumerate}
The bandwidth satisfies the standard conditions: $h \to 0$ and $nh \to \infty$ as
$n \to \infty$.

Model~\eqref{equ:general} is the most general first-order ODE constraint considered in the DE-constrained LPR framework. Several previously studied models are special cases obtained by restricting the form of $F$:
\begin{itemize}
  \item $F(x,g) = \lambda g$ gives the local exponential growth model applied to tumour growth data by \citet{ge2026differential};
  \item $F(x,g) = \lambda g^\theta$ gives the local quasi-exponential growth model studied in a companion paper;
  \item $F(x,g) = a(x)g + b(x)$ gives the general linear DE-constrained model studied in a companion paper on firebrand burning-rate data.
\end{itemize}
Consequently, the DE1-$k$ estimator developed in Section~\ref{sec:est} automatically reduces to the estimators used in those special cases when the corresponding $F$ is substituted, without any modification to the estimation procedure.

\section{Nonlinear DE-constrained Local Polynomial Estimation}
\label{sec:est}

For a given evaluation point $x_0$, the DE1-$k$ estimator of $g(x_0)$ is obtained by
minimizing the local weighted least-squares objective function
\begin{equation}
  \sum_{i=1}^n \bigl\{y_i - g_k^*(x_i;\,\alpha)\bigr\}^2 K_h(x_i - x_0)
  \label{eqn:nls}
\end{equation}
over the single local parameter $\alpha = g(x_0)$, where $g_k^*(x_i;\,\alpha)$ is the
$k$-th degree Taylor approximation of $g(x_i)$ about $x_0$ with all derivatives expressed
in terms of $\alpha$ via the ODE constraint.

\subsection{ODE-derived Taylor coefficients}
\label{sec:taylor-coef}

Since $g'(x_0) = F(x_0, g(x_0))$, all higher derivatives of $g$ at $x_0$ can be
expressed as functions of the single unknown $\alpha = g(x_0)$ by successively
differentiating $F$ with respect to $x$, using the chain rule and the ODE itself.
Specifically, let $F_x$ and $F_g$ denote the partial derivatives of $F$ with respect to
its first and second arguments, respectively. The first two derivatives are
\begin{align}
  g'(x_0)  &= F(x_0,\,\alpha), \label{eqn:d1}\\
  g''(x_0) &= F_x(x_0,\,\alpha) + F_g(x_0,\,\alpha)\,F(x_0,\,\alpha). \label{eqn:d2}
\end{align}
Higher-order derivatives $g^{(j)}(x_0)$, $j \geq 3$, are obtained analogously: each
differentiation introduces $F_x$ and $F_g$ terms which are themselves differentiated,
yielding expressions that depend only on $\alpha$, $x_0$, and the partial derivatives of
$F$. The $k$-th degree Taylor approximation is therefore
\begin{equation}
  g_k^*(x_i;\,\alpha) = \alpha + \sum_{j=1}^{k} \frac{(x_i - x_0)^j}{j!}\,
  g^{(j)}(x_0;\,\alpha),
  \label{eqn:taylor}
\end{equation}
where each $g^{(j)}(x_0;\,\alpha)$ depends on $\alpha$ alone. Crucially, only one
parameter needs to be estimated at each point $x_0$, regardless of the polynomial degree
$k$.

\subsection{Nonlinear least squares estimation}

Because $g_k^*(x_i;\,\alpha)$ is generally nonlinear in $\alpha$, minimizing
\eqref{eqn:nls} constitutes a nonlinear least squares (NLS) problem. The DE1-$k$
estimator $\widehat{g}_k(x_0)$ is defined as the solution
\begin{equation}
  \widehat{g}_k(x_0) = \arg\min_{\alpha}\,
  \sum_{i=1}^n \bigl\{y_i - g_k^*(x_i;\,\alpha)\bigr\}^2 K_h(x_i - x_0).
  \label{eqn:nlsest}
\end{equation}
In contrast to the general linear DE-constrained case, where the estimator has a
closed-form weighted least squares expression, the nonlinear case requires iterative
optimization. A reliable starting value for the NLS solver is obtained from the pilot
estimator described below.

The DE1-$k$ estimator is implemented in the
\texttt{nlODE1()} function of the \texttt{LPR4ODE} package \citep{braun2024lpr4ode}.

\subsection{Pilot estimator}
\label{sec:pilot}

The pilot estimator uses the first-degree ($k=1$) Taylor approximation
\begin{equation}
  g_1^*(x_i;\,\alpha) = \alpha + (x_i - x_0)\,F(x_0,\,\alpha),
  \label{eqn:pilot}
\end{equation}
and minimizes \eqref{eqn:nls} with this simplified model. Although less accurate than the
full DE1-$k$ estimator, the pilot provides a smooth, consistent initial curve $\widehat{g}^{\,\mathrm{pilot}}(x_0)$ over a grid of evaluation points, which serves as the
warm-start for the main NLS optimization. Warm-starting in this way substantially
improves convergence stability, particularly in regions where $F$ is strongly nonlinear.

The pilot estimator is implemented in the
\texttt{nlODE1pilot()} function of the \texttt{LPR4ODE} package \citep{braun2024lpr4ode}.

\section{Asymptotic Conditional Bias and Variance}
\label{sec:theory}

We summarize the results concerning the asymptotic conditional bias and variance of the DE1-$k$ estimator in the interior of the interval $[a,b]$ in the following two theorems.

\begin{theorem}[Asymptotic Conditional Bias]
\label{thm:nonlinear:bias}
For regression model \eqref{equ:general}, under assumptions \textnormal{(I)--(III)},
with $x_0 \in (a+h,\, b-h)$, the DE1-$k$ estimator $\widehat{g}_k(x_0)$ has asymptotic
conditional bias
\begin{equation}
  \mathrm{Bias}\bigl(\widehat{g}_k(x_0)\mid x_1,\ldots,x_n\bigr)
  = \frac{1}{(k+1)!}\,g^{(k+1)}(x_0)\,h^{k+1}\mu_{k+1} + o_p(h^{k+1}),
  \qquad k \text{ odd},
\end{equation}
and
\begin{multline}
  \mathrm{Bias}\bigl(\widehat{g}_k(x_0)\mid x_1,\ldots,x_n\bigr)
  = \left(\frac{g^{(k+2)}(x_0)}{(k+2)!}
    + \frac{g^{(k+1)}(x_0)}{(k+1)!}\frac{f'(x_0)}{f(x_0)}\right)
    h^{k+2}\mu_{k+2} \\
  + o_p(h^{k+2}), \qquad k \text{ even},
\end{multline}
where $g^{(k+1)}(x_0)$ and $g^{(k+2)}(x_0)$ are obtained by successively differentiating
$F$ as described in Section~\ref{sec:taylor-coef}.
\end{theorem}

\begin{theorem}[Asymptotic Conditional Variance]
\label{thm:nonlinear:var}
Under assumptions \textnormal{(I)--(IV)}, with $x_0 \in (a+h,\, b-h)$, the DE1-$k$
estimator $\widehat{g}_k(x_0)$ has asymptotic conditional variance
\begin{equation}
  \mathrm{Var}\bigl(\widehat{g}_k(x_0)\mid x_1,\ldots,x_n\bigr)
  = \frac{\sigma^2 R(K)}{n h f(x_0)} + o_p\!\left(\frac{1}{nh}\right).
\end{equation}
\end{theorem}

From Theorems~\ref{thm:nonlinear:bias} and~\ref{thm:nonlinear:var}, the DE1-$k$ estimator achieves the same asymptotic variance as
the classical local polynomial regression estimator of the same degree, while the bias is
determined entirely by the $(k+1)$-th or $(k+2)$-th derivative of $g$ --- quantities that
in the classical case contribute to lower-order terms. The DE-constrained approach
therefore reduces asymptotic bias without substantially increasing variance, extending to
the fully nonlinear setting the same bias--variance mechanism established for the linear
DE-constrained case.


\section{Bandwidth Selection}
\label{sec:bandwidth}

The bandwidth $h$ controls the width of the kernel window and therefore governs the
fundamental bias-variance trade-off of the DE1-$k$ estimator. From Theorems~\ref{thm:nonlinear:bias} and~\ref{thm:nonlinear:var},
decreasing $h$ reduces asymptotic bias at the cost of increased variance, while increasing
$h$ has the opposite effect. The optimal $h$ balances these two sources of error.

\subsection{AIMSE-optimal bandwidth}

Here we seek a single globally optimal $h$ that minimizes the error integrated over $[a,b]$, rather than a pointwise-optimal bandwidth.

The pointwise asymptotic mean squared error (AMSE) of $\widehat{g}_k(x_0)$ is obtained
by combining Theorems~\ref{thm:nonlinear:bias} and~\ref{thm:nonlinear:var}. For odd $k$,
\begin{equation}
  \mathrm{AMSE}\bigl(\widehat{g}_k(x_0)\bigr)
  = \frac{\mu_{k+1}^2}{\{(k+1)!\}^2}\,\bigl[g^{(k+1)}(x_0)\bigr]^2\,h^{2(k+1)}
  + \frac{\sigma^2 R(K)}{n h f(x_0)},
  \label{eqn:amse}
\end{equation}
and analogously for even $k$ using the $(k+2)$-th derivative term. Integrating
\eqref{eqn:amse} over the design density $f(x_0)$ gives the asymptotic integrated
mean squared error (AIMSE):
\begin{equation}
  \mathrm{AIMSE}(h)
  = \frac{\mu_{k+1}^2}{\{(k+1)!\}^2}\,h^{2(k+1)}
    \int_a^b \bigl[g^{(k+1)}(x)\bigr]^2 f(x)\,dx
  + \frac{\sigma^2 R(K)}{nh}\,(b-a).
  \label{eqn:aimse}
\end{equation}
Differentiating \eqref{eqn:aimse} with respect to $h$ and setting the result to zero
yields the AIMSE-optimal bandwidth
\begin{equation}
  h_{\mathrm{opt}} = \left[
    \frac{\sigma^2 R(K)\,(b-a)}
         {2(k+1)\,\dfrac{\mu_{k+1}^2}{\{(k+1)!\}^2}
          \displaystyle\int_a^b \bigl[g^{(k+1)}(x)\bigr]^2 f(x)\,dx \cdot n}
  \right]^{1/(2k+3)}.
  \label{eqn:hopt}
\end{equation}
Two unknown quantities appear in \eqref{eqn:hopt}: the error variance $\sigma^2$ and the
integrated squared derivative $\int_a^b [g^{(k+1)}(x)]^2 f(x)\,dx$. Both must be estimated from the data.

A key advantage of the DE-constrained setting is that $g^{(k+1)}(x)$ need not be
estimated directly. Since $g'(x) = F(x, g(x))$ and $F$ is known, all higher
derivatives of $g$ can be expressed entirely in terms of $g(x)$ and the partial
derivatives of $F$ via successive differentiation of the ODE. For example, for $k=1$:
\begin{equation}
  g''(x) = F_x(x, g(x)) + F_g(x, g(x))\,F(x, g(x)),
  \label{eqn:g2bw}
\end{equation}
which is a known function of $g(x)$ alone. Consequently,
$\int_a^b [g^{(k+1)}(x)]^2 f(x)\,dx$ reduces to an integral involving only $g(x)$,
which is far easier to estimate than a high-order derivative.

\subsection{Practical bandwidth estimator}

We estimate $h_{\mathrm{opt}}$ via a two-step procedure implemented in the \texttt{nlODE1bw()} function. The first step fits a
local constant smoother to obtain a pilot estimate of $g$ and an estimate of
$\sigma^2$. The second step uses the pilot estimate to approximate
$\int_a^b [g''(x)]^2 f(x)\,dx$ via the ODE constraint, avoiding the need to
estimate a second derivative directly from the data.

A local constant (Nadaraya--Watson) estimator is fitted using bandwidth $h_0 = \tfrac{1}{2}\max_i(x_{(i+1)} - x_{(i)})$, the half-maximum spacing of the
ordered design points. Let $\mathbf{S}$ denote the $n \times n$ hat matrix of this
pilot fit, with $(i,j)$-th entry
\begin{equation}
  S_{ij} = \frac{K_{h_0}(x_i - x_j)}{\sum_{\ell=1}^n K_{h_0}(x_i - x_\ell)}.
\end{equation}
The error variance is estimated by
\begin{equation}
  \widehat{\sigma}^2 = \frac{\|\mathbf{y} - \mathbf{S}\mathbf{y}\|^2}
    {n - 2\,\mathrm{tr}(\mathbf{S}) + \mathrm{tr}(\mathbf{S}^\top \mathbf{S})},
  \label{eqn:sigmahat}
\end{equation}
where the denominator is the effective degrees of freedom for residuals of the local
constant smoother \citep{fan1996local}.

For $k = 1$, the leading bias term involves $g''(x)$. From the ODE constraint
$g'(x) = F(x, g(x))$, the second derivative is
\begin{equation}
  g''(x) = F_x(x, g(x)) + F_g(x, g(x))\,F(x, g(x)),
  \label{eqn:g2}
\end{equation}
which depends on $g(x)$ only through the pilot estimates $\widehat{g}^{\,\mathrm{pilot}}(x_i)$.
The integral $\int_a^b [g''(x)]^2 f(x)\,dx$ is approximated by the Monte Carlo sum
\begin{equation}
  \widehat{I} = \frac{1}{n}\sum_{i=1}^n \bigl[\widehat{g}''(x_i)\bigr]^2,
  \label{eqn:Ihat}
\end{equation}
where $\widehat{g}''(x_i)$ is \eqref{eqn:g2} evaluated using the pilot estimates of $g(x)$.

Substituting $\widehat{\sigma}^2$ and $\widehat{I}$ into a simplified form of
\eqref{eqn:hopt} --- absorbing the kernel constants $\mu_{k+1}$, $\{(k+1)!\}$, and
$2(2k+1)$ into the rule-of-thumb --- gives the data-driven bandwidth
\begin{equation}
  \widehat{h} = \left(
    \frac{R(K)\,\widehat{\sigma}^2\,(b-a)}{n\,\widehat{I}}
  \right)^{1/5},
  \label{eqn:hhat}
\end{equation}
where $R(K) = 1/(2\sqrt{\pi})$ for the Gaussian kernel. The exponent $1/5$ is the
standard optimal rate for local linear regression \citep{fan1996local}.

\subsection{Practical recommendations}

In practice, the bandwidth is selected as follows. The direct plug-in bandwidth selector
\texttt{dpill()} from the \texttt{KernSmooth} package \citep{kernsmooth} provides a
quick, reliable starting value for the pilot step. The pilot estimator \texttt{nlODE1pilot()}
is then run with this bandwidth to obtain $\widehat{g}^{\,\mathrm{pilot}}$, which is used
both as a warm-start for the main NLS and as input to \eqref{eqn:g2}--\eqref{eqn:Ihat}.
The main DE1-$k$ estimator \texttt{nlODE1()} is run with bandwidth $\widehat{h}$ from
\eqref{eqn:hhat}, or with $2 \times \texttt{dpill()}$ as a simpler alternative when the
ODE is strongly nonlinear and variance stability is the primary concern.

\section{Simulation Study}
\label{sec:sim}

We evaluate the DE1-$k$ estimator through a simulation study based on logistic
population growth, a canonical nonlinear ODE model, and compare it against the
PCODE method of \citet{ramsay2007parameter}, a representative functional data
analysis (FDA) approach.

\subsection{Data-generating model}

The true mean function $g(x)$ satisfies the logistic growth equation
\begin{equation}
  g'(x) = \theta\, g(x)\!\left(1 - \frac{g(x)}{10}\right),
  \quad g(0) = 0.1,\quad \theta = 0.1,
  \label{eqn:logistic}
\end{equation}
on the interval $[0, 100]$. The solution of \eqref{eqn:logistic} is the standard
logistic curve with carrying capacity 10 and intrinsic growth rate $\theta = 0.1$.
Observations are generated as
\begin{equation}
  y_i = g(x_i) + \varepsilon_i, \quad \varepsilon_i \sim N(0,\, \sigma^2),
  \quad i = 1, \ldots, n,
  \label{eqn:sim:model}
\end{equation}
with $n = 101$, noise level $\sigma = 0.5$, and design points $x_i$ drawn independently
from $\mathrm{Uniform}[0, 100]$. The true curve $g(x)$ is computed by numerically solving \eqref{eqn:logistic} using
the \texttt{ode()} solver from the \texttt{deSolve} package
\citep{soetaert2010solving}.

\subsection{Methods compared}

Three estimation methods, the DE1-3 estimator, PCODE, and local linear regression, are applied to each simulated dataset.

The DE1-$k$ estimator with $k = 3$ (third-degree local polynomial) is fitted using
\texttt{nlODE1()} \citep{braun2024lpr4ode} with the ODE constraint $F = 0.1\,g(1 -
g/10)$ treated as fully known. The bandwidth is set to $h = 2\,\tilde{h}$, where
$\tilde{h} = \texttt{dpill}(x, y)$ is the direct plug-in selector from the
\texttt{KernSmooth} package \citep{kernsmooth}. A pilot fit from \texttt{nlODE1pilot()} \citep{braun2024lpr4ode} with bandwidth $\tilde{h}$ provides
warm-start values for the main NLS optimization. The estimator is evaluated on a
grid of 401 equally spaced points over $[0, 100]$.

The parameter cascade method of \citet{ramsay2007parameter} is
used as the principal comparator, implemented in the
\texttt{pCODE} package \citep{wang2022pcode}. The method
represents the solution curve in a B-spline basis and
estimates the structural parameters of the ODE by a nested
optimization in which the spline coefficients are treated as
nuisance parameters, profiled out at an inner level, while the
structural parameters are estimated at an outer level. The two
levels are linked by a penalty that measures the extent to
which the fitted spline fails to satisfy the differential
equation, so the smoothing parameter governs how strictly the
ODE is enforced rather than how smooth the curve is.

In the present study the basis consists of 23 B-spline
functions of polynomial order 4, formed from 21 equally spaced
knots on $[0, 100]$, and the ODE-fidelity penalty is set to
$\lambda = 100$. The structural parameter is the intrinsic
growth rate $\theta$ of \eqref{eqn:logistic}, initialized at
$\theta_0 = 0.5$ and estimated jointly with the spline
coefficients; the carrying capacity is held fixed at its true
value of $10$, so that PCODE and DE1-3 are supplied with the
same structural information apart from $\theta$ itself. The
fitted curve is reconstructed from the estimated coefficients
and basis, and evaluated on the same grid of 401 points used
for DE1-3.

Two features of this configuration matter for interpreting the
results of Section~\ref{sec:sim:results}. First, the
comparison is deliberately favourable to PCODE: it is fitted
under the correct parametric family and estimates $\theta$
from the data, whereas DE1-3 imposes the ODE as a known
constraint without assuming any parametric form for $g$. The
two estimators are therefore not competing on equal terms, and
the relevant question is how much of PCODE's accuracy the
nonparametric estimator recovers. Second, the outer
optimization over $\theta$ is a nonconvex problem whose
solution depends on the initial value; this is the source of
the convergence failures and spurious negative estimates
reported at large $\theta$ in
Section~\ref{subsec:sensitivity}, and it has no analogue in
DE1-3, where no structural parameter is estimated.

Classical local linear regression is fitted using \texttt{locpoly()} from \texttt{KernSmooth} \citep{kernsmooth} with bandwidth $\tilde{h} =
\texttt{dpill}(x, y)$. This method ignores the ODE constraint entirely and serves
as a fully nonparametric baseline.

\subsection{Results}
\label{sec:sim:results}

\begin{figure}[htbp]
  \centering
  \includegraphics[width=\textwidth]{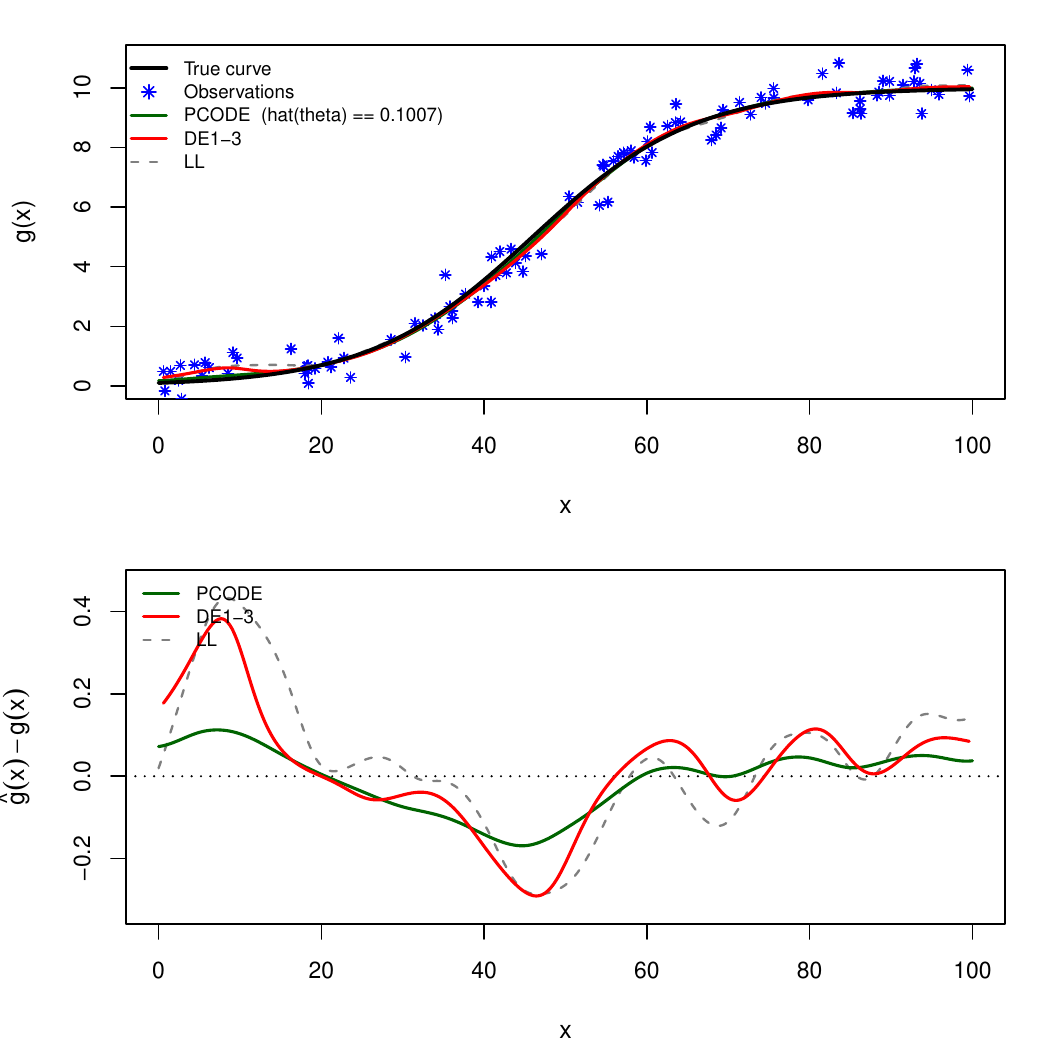}
  \caption{Representative simulation run for the logistic growth model \eqref{eqn:logistic} with $\theta = 0.1$, $n = 101$, $\sigma = 0.5$.
           \textit{Top panel}: solid black line: true curve $g(x)$;
           blue stars: observations;
           green line: PCODE fit ($\hat{\theta} = 0.1007$);
           red line: DE1-3 fit;
           dashed grey line: classical local linear regression (LPR).
           \textit{Bottom panel}: pointwise residuals $\widehat{g}(x) - g(x)$
           for each method; dotted horizontal line marks zero.
           Bandwidth for DE1-3: $h = 2\,\tilde{h}$ where
           $\tilde{h} = \texttt{dpill}(x, y)$.}
  \label{fig:sim:logistic}
\end{figure}

Figure~\ref{fig:sim:logistic} shows a representative single realization of the simulation. The true logistic curve (solid black), the noisy observations (blue stars), the PCODE fit (green), and the DE1-3 fit (red) are displayed together. The
DE1-3 estimator closely tracks the true curve throughout the range, including the inflection region near $x = 23$. The PCODE fit is also close to the true curve, reflecting the advantage of its parametric structure, but requires estimating
$\theta$ whereas DE1-3 treats the ODE as known.

Table~\ref{tab:sim:imse} summarizes the average MSE over all 500 replications. Although PCODE achieves the lowest average MSE of $0.00753$ (SE $= 0.000304$), this comparison is inherently asymmetric: PCODE is fitted under the correct parametric model and estimates the structural parameter $\theta$ jointly with the
spline curve, whereas DE1-3 is a fully nonparametric estimator that uses only the ODE constraint without assuming any parametric form for $g$. The DE1-3 estimator achieves an average MSE of $0.01383$ (SE $= 0.000319$), which represents a reduction of approximately $44\%$ relative to classical local linear regression
(average MSE $= 0.02458$, SE $= 0.000517$), demonstrating that the ODE constraint provides substantial information even within a nonparametric framework. Moreover, in a small number of replications PCODE converged to a spurious negative value of
$\hat{\theta}$, indicating sensitivity to the choice of initial value in the optimizer. The DE1-3 estimator, requiring no estimation of $\theta$, is immune to this form of instability. In settings where the structural parameter is difficult
to identify, or where the parametric ODE model is only approximately correct, the nonparametric DE1-3 estimator is expected to be more competitive.

The top panel of Figure~\ref{fig:sim:logistic} shows that all three methods track the true logistic curve closely, but differences are subtle at this scale. The bottom panel displays the pointwise residuals $\widehat{g}(x) - g(x)$
for each method, revealing the structure of the errors more clearly.

In the slow-growth region $x \in [0, 20]$, classical LPR exhibits a large positive spike, being pulled upward by noisy observations in the absence of any structural constraint. Both DE1-3 and PCODE remain much closer to zero in this region, demonstrating that the ODE constraint effectively anchors the fit where the data
alone provide limited information.

In the inflection region $x \in [35, 55]$, DE1-3 shows a negative dip of approximately $-0.3$, reflecting mild oversmoothing bias where $g$ changes most rapidly. This is the hardest region for a fixed-bandwidth estimator, and a locally adaptive bandwidth would be expected to reduce this error. PCODE's dip in the same
region is smaller (approximately $-0.2$), owing to the stabilizing effect of its global parametric constraint on $\theta$. LPR oscillates more erratically throughout the range, consistent with its higher average MSE in Table~\ref{tab:sim:imse}.

In the plateau region $x \in [70, 100]$, all three methods converge to similar residuals near zero, since the curve is nearly flat and straightforward to estimate without structural information.

Overall, the residual panel confirms that the ODE constraint is most valuable in regions where the curve changes slowly or where data are sparse relative to the rate of change, precisely the settings where classical local polynomial regression is
most prone to oversmoothing or being misled by noise.

\begin{table}[htbp]
  \centering
  \begin{tabular}{lcc}
    \toprule
    Method & Average IMSE & Std.\ Error \\
    \midrule
    LL & 0.02458            & (0.000517)  \\
    DE1-3              & 0.01383            & (0.000319)  \\
    PCODE              & 0.00753            & (0.000304)  \\
    \bottomrule
  \end{tabular}
    \caption{Average integrated mean squared error (IMSE) over $N = 500$ Monte Carlo replications for the logistic growth simulation. Design: $x_i \sim \mathrm{Uniform}[0,100]$, $n = 101$, $\sigma = 0.5$. \textit{Note}: MSE is computed as the average squared difference between the fitted and true curves at the observed design points. Methods are ordered by decreasing MSE.}
     \label{tab:sim:imse}
\end{table}

\subsection{Sensitivity analysis: effect of growth rate $\theta$}
\label{subsec:sensitivity}

The simulation study above used a fixed growth rate $\theta = 0.1$. To assess how the relative performance of the three methods depends on the shape of the true curve, we repeat the simulation for $\theta \in \{0.05, 0.10,
0.20, 0.50\}$, keeping all other settings identical ($n = 101$, $\sigma = 0.5$, $N = 500$ replications). Larger values of $\theta$ produce a steeper S-curve with a sharper inflection, making the estimation problem progressively harder for methods
that do not exploit the ODE structure.

Table~\ref{tab:sens} reports the average MSE and the bias of $\hat{\theta}$ from PCODE for each value of $\theta$.

\begin{table}[htbp]
  \centering
  \begin{tabular}{ccccc}
    \toprule
    $\theta$ & LL MSE & DE1-3 MSE & PCODE MSE
             & PCODE bias($\hat{\theta}$) \\
    \midrule
    0.05 & 0.01694 & 0.01199 & 0.00765 & $-0.00014$ \\
         & (0.000472) & (0.000419) & (0.000241) & \\
    0.10 & 0.02350 & 0.01410 & 0.00635 & $-0.00026$ \\
         & (0.000487) & (0.000382) & (0.000221) & \\
    0.20 & 0.03125 & 0.01292 & 0.00520 & $-0.00027$ \\
         & (0.000526) & (0.000305) & (0.000203) & \\
    0.50 & 0.04960 & 0.00857 & 0.02209 & $-0.07764$ \\
         & (0.000655) & (0.000253) & (0.001275) & \\
    \bottomrule
  \end{tabular}
    \caption{Sensitivity analysis: average MSE over $N = 500$ replications for
           $\theta \in \{0.05, 0.10, 0.20, 0.50\}$.
           Design: $x_i \sim \mathrm{Uniform}[0,100]$, $n = 101$, $\sigma = 0.5$.
           Standard errors in parentheses.
  \textit{Note}: DE1-3 NAs: 0, 0, 0, 5 for
  $\theta = 0.05, 0.10, 0.20, 0.50$ respectively.}
    \label{tab:sens}
\end{table}

\begin{figure}[htbp]
  \centering
  \includegraphics[width=\textwidth]{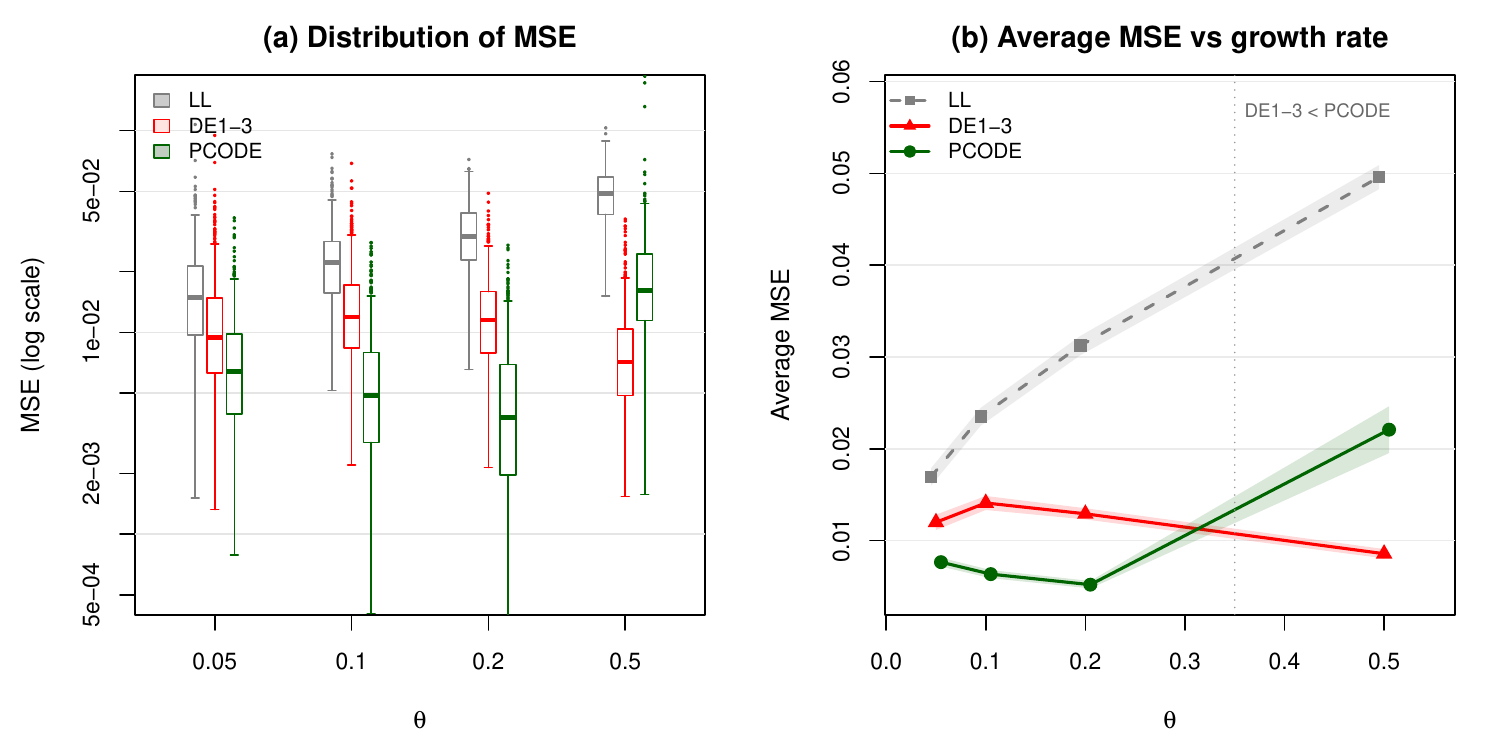}
  \caption{Sensitivity analysis for the logistic growth simulation
           ($n = 101$, $\sigma = 0.5$, $N = 500$ replications).
           \textit{Left panel~(a)}: distribution of MSE on a log scale for each method and $\theta$ value; boxes show interquartile range, whiskers extend to 1.5$\times$IQR, and outliers are shown as dots.
           \textit{Right panel~(b)}: average MSE versus $\theta$ with $\pm 2$ standard error bands; the vertical dotted line marks the approximate crossover point beyond which DE1-3 outperforms PCODE. Grey dashed: LL; red solid: DE1-3; green solid: PCODE.}
  \label{fig:sens}
\end{figure}

Several patterns emerge from Table~\ref{tab:sens}, which are further illustrated
in Figure~\ref{fig:sens}.

Table~\ref{tab:sens} and Figure~\ref{fig:sens} show that DE1-3 outperforms conventional local linear regression at every value of $\theta$, with the gap widening as the growth rate increases.
A steeper logistic curve changes more rapidly over $[0, 100]$, so a kernel smoother with no knowledge of the ODE structure becomes increasingly unreliable, while DE1-3 incorporates the shape of $F$ directly into the approximation at every evaluation point.

The comparison with PCODE is more nuanced. For $\theta \leq 0.20$, PCODE outperforms
DE1-3, which is expected since PCODE fits the correct parametric model and estimates
$\theta$ jointly with the curve. At $\theta = 0.50$, however, PCODE's average MSE
rises to $0.02209$ while DE1-3 achieves $0.00857$. The PCODE estimate $\hat\theta$
shows a bias of $-0.078$ at this value, $15.6\%$ of the true value, and the
optimizer failed to converge in 5 of 500 replications. The steep inflection of the
logistic curve at $\theta = 0.50$ apparently creates a difficult optimization landscape
for the parameter cascade method. Since DE1-3 treats $\theta$ as known, it is
unaffected by this instability.

Overall, the results suggest a trade-off: PCODE is more accurate when the structural
parameter is reliably estimable, but DE1-3 is more robust when it is not. A
nonparametric approach like DE1-3 sidesteps the identifiability issue altogether, which
may be an advantage in settings where the experimental design or noise level makes
parametric estimation unreliable.

\section{Discussion}
\label{sec:discussion}

The theoretical and simulation results in this paper are mutually consistent. Theorems~\ref{thm:nonlinear:bias} and~\ref{thm:nonlinear:var} show that the DE1-$k$ estimator matches the asymptotic variance of classical local polynomial regression of the same degree while reducing bias to a term governed by the $(k+1)$-th or $(k+2)$-th derivative of $g$, exactly the mechanism established previously for the exponential and general linear special cases. The logistic growth simulation confirms this pattern in finite samples and, more importantly, characterizes where the ODE constraint is most valuable relative to a competing parametric approach: DE1-3 improves on classical local linear regression at every growth rate considered, with the advantage increasing as the underlying curve steepens, and it becomes more accurate than PCODE precisely in the regime, large $\theta$, where PCODE's nonconvex outer optimization over the structural parameter becomes unreliable.

This complementary pattern is, we think, the central practical message of the comparison. PCODE and other parameter cascading methods are highly effective when the parametric form of the ODE is exactly correct and its structural parameters are well identified by the data; in that regime, the extra information contained in a correctly specified parametric model is difficult for any nonparametric method to match, and Table~\ref{tab:sim:imse} shows PCODE winning comfortably at $\theta = 0.1$. What DE-constrained LPR offers is a different guarantee: because it never estimates a structural parameter, it cannot suffer the convergence failures or biased parameter estimates that degrade PCODE's performance under a difficult identification problem, and its accuracy therefore degrades more gracefully as the estimation problem becomes harder. This suggests a natural role for DE-constrained LPR as either a robust default when structural identifiability is in doubt, or as a diagnostic check on a parameter cascading fit: substantial disagreement between the two methods would flag a case where the parametric estimate of $\theta$ deserves scrutiny.

The unification noted in Section~\ref{sec:model}, that the exponential, quasi-exponential, and general linear DE-constrained models are all special cases of the general nonlinear framework developed here, is more than a formal observation. Because the DE1-$k$ estimator is implemented generically in terms of symbolic differentiation of $F$, the same \texttt{nlODE1()} routine used for the logistic growth model applies without modification to any first-order ODE constraint, including those studied in our companion papers on tumour growth and firebrand burning rate. This makes the general nonlinear framework a natural common implementation layer for the DE-constrained LPR program as a whole, rather than a fourth, disconnected special case.

Several limitations point to directions for future work. First, as noted in Theorem~\ref{thm:nonlinear:bias}, the asymptotic theory is pointwise; the AIMSE-optimal bandwidth of Section~\ref{sec:bandwidth} addresses this by integrating over the design density, but a fully locally adaptive bandwidth, varying $h$ with the local curvature of $F$, would likely reduce the oversmoothing bias observed near the inflection point of the logistic curve in Section~\ref{sec:sim:results}. Second, the practical bandwidth estimator of Section~\ref{sec:bandwidth} relies on a pilot local-constant fit whose own error propagates into $\widehat{h}$; characterizing this propagation formally, as opposed to relying on simulation evidence of good practical performance, is left for future work. Third, the simulation study considers a single canonical nonlinear ODE (logistic growth); extending the sensitivity analysis to other nonlinear structures, such as Michaelis--Menten kinetics or SIR-type compartment models, would help establish how general the observed crossover pattern between DE1-3 and PCODE really is.

\section{Conclusions}
\label{sec:conclusions}

This paper developed differential equation-constrained local polynomial regression for the fully general nonlinear first-order ODE constraint $g'(x) = F(x, g(x))$, extending the DE-constrained LPR framework beyond the exponential and general linear special cases studied previously. The DE1-$k$ estimator is constructed by expressing all Taylor coefficients of $g$ at a local center $x_0$ as functions of the single unknown $\alpha = g(x_0)$, using successive symbolic differentiation of $F$ via the chain rule; the resulting one-dimensional nonlinear least squares problem is solved iteratively with warm starts from a pilot estimator, and only one local parameter is estimated at each evaluation point regardless of the polynomial degree $k$.

We established the asymptotic conditional bias and variance of $\widehat{g}_k(x_0)$: the variance matches that of classical local polynomial regression of the same degree, while the bias is governed by derivatives of order $k+1$ or $k+2$ of $g$, which the classical estimator would only reach at lower order. We derived the AIMSE-optimal bandwidth analytically and proposed a practical two-step estimator that exploits the ODE structure to avoid direct estimation of a high-order derivative.

A simulation study based on logistic population growth compared the DE1-3 estimator against the parameter cascading method PCODE and classical local linear regression. DE1-3 consistently outperformed local linear regression, with the advantage growing from $29\%$ at $\theta = 0.05$ to $83\%$ at $\theta = 0.50$. PCODE achieved a lower MSE than DE1-3 at small and moderate growth rates, where its correctly specified parametric model and jointly estimated structural parameter give it an advantage, but at the steepest growth rate considered ($\theta = 0.50$) DE1-3 outperformed PCODE, whose optimizer produced a large bias in $\hat\theta$ and failed to converge in a nontrivial fraction of replications. The nonparametric DE1-3 estimator, which estimates no structural parameter, is immune to this form of instability.

Together with the exponential and general linear special cases studied in companion work, these results position DE-constrained LPR as a broadly applicable and computationally uniform framework for nonparametric estimation under mechanistic ODE constraints, and identify parameter identifiability, rather than curve smoothness alone, as the key factor governing when a nonparametric, structure-respecting estimator such as DE1-$k$ is likely to outperform a parametric ODE-fitting method such as PCODE.

\section*{Acknowledgements}

The \texttt{LPR4ODE} package used throughout this paper was developed by W.\ J.\ Braun. This research has been supported in part by a grant from the Natural Sciences and Engineering Research Council of Canada (NSERC).

\bibliographystyle{apalike}
\bibliography{references}

\end{document}